\documentclass[10pt]{article}
\usepackage{graphicx}
\usepackage{amssymb}
\usepackage{epstopdf}
\title{A stochastic model for wound healing}
\author{Thomas Callaghan$^1$, Evgeniy Khain$^{2,3}$,\\ Leonard M. Sander$^{2,3}$, and Robert M. Ziff$^{2,4}$ \\
$^1$ School of Mathematics, Georgia Institute of Technology \\
$^2$ Michigan Center for Theoretical Physics\\
$^3$ Department of Physics, University of Michigan\\
$^4$ Department of Chemical Engineering, University of Michigan }
\begin{document}
\maketitle
\begin{abstract}
We present a discrete stochastic model which represents many of the salient features of the biological process of wound healing. The model describes fronts of cells invading a wound.  We have numerical results in one and two dimensions. In one dimension we can give analytic results for the front speed as a power series expansion in a parameter, $p$, that gives the relative size of proliferation and diffusion processes for the invading cells. In two dimensions the model becomes the Eden model for  $p \approx 1$. In both one and two dimensions for small $p$, front propagation  for this model should approach that of the Fisher-Kolmogorov equation. However, as in other cases, this discrete model approaches Fisher-Kolmogorov behavior  slowly. 
\end{abstract} 
\section{Introduction}
The biology of wound healing is fairly well understood \cite{murray02}.  A simplified version of the process may be given as follows: a layer of undamaged cells is usually  quiescent, so that the birth rate of cells matches the death rate, and both are quite small. When a wound is suffered, there is a rapid signal the wakes the cells up -- perhaps a pulse of ATP or a calcium wave. Cells at the edge of the wound become more mobile, and also enhance their  proliferation rate. (Otherwise the healed layer would not have the right density.)  A typical experiment to study this process consists in plating suitable cells (e.g. epithelial cells) on a substrate so that they form a confluent monolayer. Then a scratch is made in the layer, and the process of filling in the scratch is studied. For example, the speed of advance of the invading cells, $v$, is easily measured.

There have been many modeling studies of wound healing \cite{sheardown96,sheratt02,maini04,walker04a}. In  many cases (\cite{walker04a} is an exception) the process is studied using some variant of the Fisher-Kolmogorov (FK) equation \cite{fisher37,kolmogorov37}. This is an obvious model to use. It builds in  diffusion with diffusion constant $D$ and proliferation with growth rate $k$ (related to inverse doubling time). It also shuts off growth for the confluent layer at density $c_o$. 
\begin{equation}
 \partial c/\partial t = D \nabla^2 c + kc (1-c/c_o) 
\label{FKeqn}
\end{equation}
The justification for using a continuum equation for a cellular process relies on the common experience that coarse-graining is reasonable for dynamic processes involving a large number of agents. In this particular case, we expect that the FK equation should be useful if  the characteristic length of the pattern predicted by Eq.~(\ref{FKeqn}) is much larger that the size of a cell. 

However, it is well known \cite{brunet97,kessler98,moro03} that coarse-graining the FK equation has many pitfalls even in this limit, and that the transition to the continuum limit is often very slow.  This motivates the present investigation: we present a discrete stochastic model for wound healing, and study it in various limiting regimes. It is quite similar to a model previously introduced and studied for flame-front propagation \cite{bramson86,kerstein86}. Thus, our results and methods should be of interest beyond the explicit biological context. We will give new numerical and analytical results, and show how, in one and two dimensions, our model aproaches the FK limit. We will show that in the biologically relevant regime there are corrections to FK due to discreteness.  

\section{Formulation of the model and known properties}
Consider a set of sites that form a linear or square lattice, corresponding to one or two dimensional `tissues'. We allow each site to be occupied by zero or one cells. Our initial configuration is an occupied half space: if $i$ labels the $x$ coordinates of the sites, then we have all sites with $i \le 0$ occupied. The dynamical rules are as follows: we choose a parameter $p$ which specifies the proliferation rate of the cells. Then at any time step we choose a cell at random, and an adjacent site at random as a target for diffusion or proliferation. E.g., in 1d if we choose a cell at site $i$, we also pick site $i + 1$ or $i-1$ as a target. If the target site is empty, with probability $p$ we put a new cell at the target, and with probability $q=1-p$ we move the chosen cell to the target.  If the target site is filled, we do nothing. 

These are examples of the elementary  processes allowed:
\begin{itemize}
  \item (...1111000...) $\to$ (...1111100...); probability $p$
  \item (...1111000...) $\to$ (...1110100..); probability $q$
\end{itemize}

As time advances cells appear for $i>0$. These form a front or chemical wave. We will examine the speed, $v(p)$, and front width, $w(p)$ for the invading cells. Precise definitions for these quantities will be given below.

An essentially identical model was devised by Kerstein \cite{kerstein86} to describe flame-front propagation. He studied it numerically in 1d, and Bramson \emph{et al.} \cite{bramson86} found some analytic results, also in 1d. In their formulation there is a parameter $\gamma$ which may be identified as $(1-p)/p$ in our notation. Also, in their model the time unit is different from ours by a factor $1+\gamma$. If $V(\gamma)$ denotes the front speed in the Kerstein model, we have:
\begin{equation}
\label{convertv}
v(p) = V(\gamma)/(1+\gamma).
\end{equation} 

In \cite{bramson86} there are two exact results. In our notation these are:
\begin{itemize}
  \item $v(p) \to 1/2 + {\cal O} (q^2)$ as $p \to 1$,
  \item $ v(p) \to  \sqrt{2p}$ as $p \to 0$.
\end{itemize} 
The first of these two is obvious. In the limit $p=1$ there is no diffusion, only proliferation, and the half space advances with no vacancies. The only process allowed is to choose the leading cell at site $i$ and proliferate at site $i+1$. Since half the moves are wasted by choosing as a target the filled site at $i-1$ the front speed is 1/2. The lack of a term linear in $q$ will be derived below. 

The second result may be understood by comparison with Eq.~(\ref{FKeqn}). Consider the coarse-grained limit of our discrete model using the lattice constant as the unit of space, and a computer time step as the time unit. It is elementary to see that the diffusion coefficient, $D$, is   1/2. Now consider a collection  of cells distant from one another with concentration $c$. In unit time the number will increase to $(1+p)c$. By integrating Eq.~(\ref{FKeqn}) over space in the low density limit, we see that we must identify $k=p$. The front velocity given by  Eq.~(\ref{FKeqn}) is well known for bounded initial conditions \cite{panja04}:
\begin{equation}
\label{continuumv}
  v = 2\sqrt{Dk} = \sqrt{2p}.
\end{equation}
Kerstein \cite{kerstein86} verified both limits numerically. We will extend these one dimensional results below both numerically and analytically, and also investigate the two-dimensional case.

\begin{figure}
\begin{center}
\includegraphics[width=3.5in]{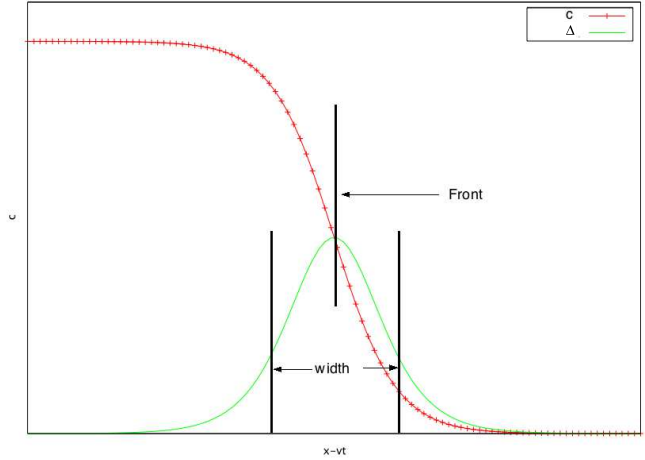}
\caption{Sketch of traveling wave solution to the FK equation. The front position and width can be defined as shown. $\Delta$ is the negative derivative of $c$, see text.}
\label{sketch}
\end{center}
\end{figure}

We note for future reference that the solutions of Eq.~(\ref{FKeqn}) generate an interface with an intrinsic width (see Figure~(\ref{sketch})) given by:
\begin{equation}
\label{continuumw}
w = \sqrt{D/k} \propto 1/\sqrt{p}.
\end{equation}
Previous authors have not discussed the front width, but, as we will see, it is relevant to a biological interpretation of the results.

\section{Numerical results}
\subsection{Defining the front}
The solution to Eq.~(\ref{FKeqn}) is a traveling front of the general form shown in the sketch in Figure~\ref{sketch}. Our data for the discrete model is the form of occupancies of sites as a function of time. We present here  a useful way to analyze such data that allows easy comparison to continuum theories. 

We start by defining the occupancy of a given column of our numerical data, $P(i)$. In 1d this is simply 1 or 0, depending on whether site $i$ is occupied. In 2d it is the average occupancy of column $i$, that is, the number of occupied sites with first coordinate $i$ divided by the width of the system (the total number of such sites) which we denote by $L$. We also define the negative of the discrete derivative of $P$; it  is localized near the interface:
\begin{equation}
\label{defD}
\Delta(i) = P(i) - P(i+1). 
\end{equation}
Note that at long times we certainly have $P(0)=1$, and for large enough $i, P(i)=0$. Thus:
\begin{equation}
\label{sumD}
\sum_{i=0}^{\infty} \Delta(i) =1.
\end{equation}
That is, we can use $\Delta$ as a weight function to define averages. We put:
\begin{eqnarray}
\langle i \rangle  & = & \sum_{i=0}^\infty i\Delta(i) =  \sum_{i=1}^\infty P(i) = n_p \nonumber \\
\langle i^2 \rangle & = &  \sum_{i=0}^\infty i^2 \Delta(i) = \sum_{i=1}^\infty (2i-1)P(i), \cdots
\end{eqnarray}
Here, $n_p$ is the number of particles for $i \ge 1$ in 1d, or that number divided by $L$ in 2d. That is, we get the position of the front by the total mass of created particles. 

The front speed is defined as $$v= \lim_{t \to \infty} \langle i \rangle/t.$$ The front width is given by $$w=\sqrt{\langle i^2 \rangle - \langle i \rangle^2}.$$ Other moments of the distribution can be defined similarly.

\subsection{One dimension}
\begin{figure}
\begin{center}
\includegraphics[width=3.5in]{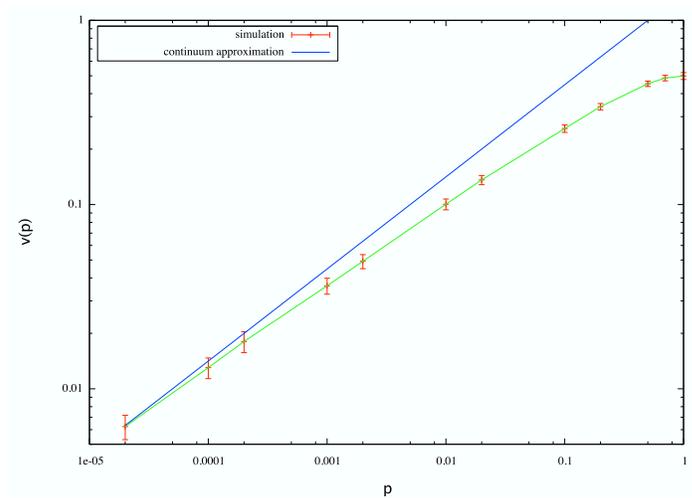}
\caption{The front speed, $v(p)$ in one dimension. The numerical simulations are averaged over 50 realizations for large $p$ and up to 1000 for the smallest $p$ to give the errorbars. Upper line is the continuum approximation, Eq.~(\ref{continuumv}). }
\label{v(p)}
\end{center}
\end{figure}

\begin{figure}
\begin{center}
\includegraphics[width=3.5in]{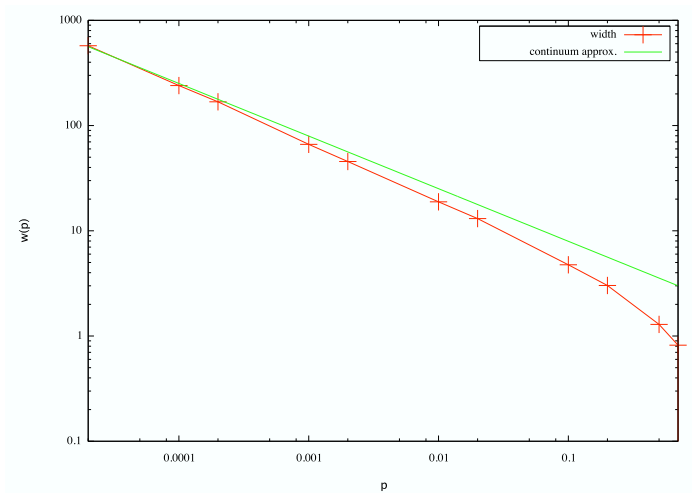}
\caption{The front width, $w(p)$ from the same simulations as Fig.~(\ref{v(p)}). The width is arbitrary up to a numerical factor. Hence the continuum approximation (upper line),
 Eq.~(\ref{continuumw}) is multiplied by a fitting factor.  }
\label{w(p)}
\end{center}
\end{figure}

The results of our simulations are shown in Figures~(\ref{v(p)}) and (\ref{w(p)}). The front speeds were found by fitting $\langle i \rangle$ to $vt$. It is remarkable that the front speed is quite well defined even for very small $p$. Of course, for $p=0$ the speed is not defined at all. 

The convergence to the continuum predictions is quite evident in the figures. Note that the prediction for $v(p)$ does not contain an adjustable constant, so the agreement is quite remarkable.  However, following the work of  \cite{moro03,brunet97} we would expect the corrections to the continuum prediction, Eq.~(\ref{continuumv}), to follow:
\begin{equation}
\label{moroformula}
v(p) = \sqrt{2p} - A/\ln^2(p),
\end{equation}
where $A$ is a factor of order unity. In fact, this expression does not fit our results. Rather, the correction to the continuum formula is more like a power law with a power near 2/3. 

\subsection{Two dimensions}
\begin{figure}
\begin{center}
\includegraphics[width=3.5in]{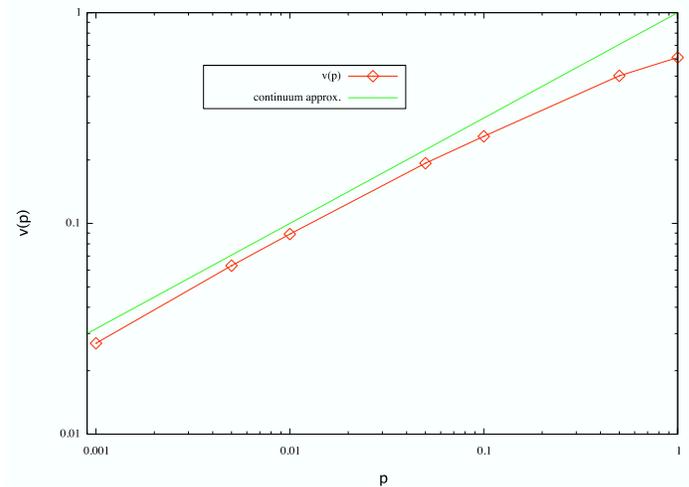}
\caption{The front speed in two dimensions. The continuum approximation is from Eq.~(\ref{continuumv2d}).}
\label{v2d}
\end{center}
\end{figure}

Our numerical results for $v(p)$ in 2d are given in Figure~\ref{v2d}. Note that as $p \to 1, v(p) > 0.5$.  Analysis of the processes the contribute to front motion in 2d is more complex than in 1d where $v(1)=0.5$ is an exact result. In particular the front will always be rough (see below) so that particles behind the leading particle will not be blocked from advancing.

For $p\ll 1$ we expect that we should converge to the result of the FK equation, namely that $v(p) \propto \sqrt{p}$. However, for 2d we have no convincing \emph{a priori} estimate of the prefactor. Following the treatment above, we might proceed by noting that in 2d $D=1/4$ for the discrete model. We then have:
\begin{equation}
\label{continuumv2d}
v(p) \approx \sqrt{p}.
\end{equation}
As we can see from the figure, this is a reasonable estimate for small $p$. 

\begin{figure}
\begin{center}
\includegraphics[width=3.5in]{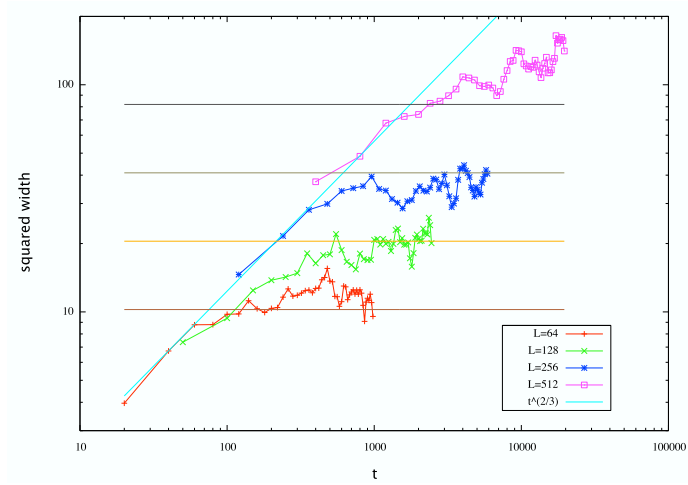}
\caption{Scaling of $w^2(p=1)$ with time. From bottom to top, $L=64, 128, 256, 512$. Also
shown are lines giving the expected scaling for early times, $w^2 \propto t^{2/3}$, and late times, $w^2 \propto L$. }
\label{edenscaling}
\end{center}
\end{figure}

\begin{figure}
\begin{center}
\includegraphics[width=3.5in]{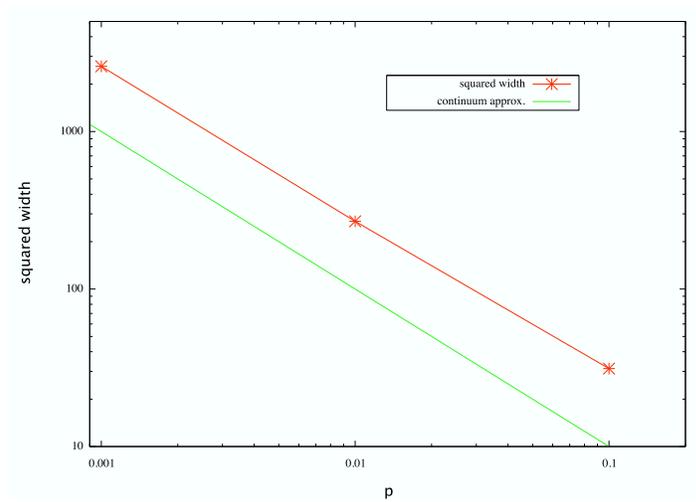}
\caption{The saturated value of $w^2$ in 2d. Also shown is the continuum approximation, $w^2 \propto 1/p$.}
\label{w2d}
\end{center}
\end{figure}

In two dimensions the width of the interface is a more complicated object than in one dimension \cite{panja04}. The reason for this is that in the presence of fluctuations the front can do two different things: it can spread so that it has an intrinsic width (as in 1d) by having a reduced density in the interface region, but also it can \emph{wander}. Indeed,  for $p=1$ wandering is the only effect possible. (Recall that in 1d $w(p=1)=0$.) In fact, in the large $p$ regime this model is identical to the Eden model \cite{eden61} where perimeter sites all grow with equal probability. 

The phenomenology of the Eden model is well understood \cite{panja04}. The wandering of the interface is time-dependent and obeys (in our system of units):
\begin{eqnarray}
w & \propto & t^{1/3} \quad t \ll L^{3/2} \nonumber \\
 & \propto  & L^{1/2} \quad t \gg L^{3/2}.
\label{edenscalingeq} 
\end{eqnarray}
This is indeed the case here: see Figure~\ref{edenscaling}. The agreement with the Eq.~(\ref{edenscalingeq}) is reasonable. We have verified that the scaling behavior given in Eq.~(\ref{edenscalingeq}) persists down to $p=0.5$.

However, as $p$ decreases the intrinsic width grows rapidly. As soon as the intrinsic width exceeds the saturated width from wandering (the second line of Eq.~(\ref{edenscalingeq})) we will loose the power-law time dependence of $w$. In Figure~\ref{w2d} we show the saturated width for a range of $p$.

\section{Results for  $p \approx 1$ in one dimension}
For $p \approx 1$ the dominant process is proliferation. For $p=1$ this gives rise to a simple configuration as we have mentioned above:  all sites behind the front are occupied, and the front advances because the leading cell proliferates. For $q \ll 1$ there is a small probability $q/2$ of creating a configuration with a  `hole'. Because the model is very simple we can use this observation to work out the power series expansion of $v(q)$.
  
\subsection{Exact solution of model for states with one hole}  
Suppose we consider only states with zero holes or one hole at any position. We expect these to be the dominant configurations small $q$.
Define the states:
\begin{itemize}
\item $|0\rangle = (...11111000...)$
\item  $|1\rangle= (...11101000...)$
\item $|2\rangle= (...11011000...)$
\item  $|3\rangle= (...10111000...)$, etc.
\end{itemize}
%The cells proliferate with probability $p$, or  move with  probability $q$, assuming the target site is empty. 
We allow transitions only between these states.
% with at most one hole --- we do not allow transitions to other states.
The transitions and their associated probabilities 
 $W_{ij} \equiv W(|i\rangle \to |j\rangle)$ are:
\begin{eqnarray}
W_{00} =p/2 \quad W_{01} = q/2 \nonumber \\ 
W_{10} = (1+p)/2 \quad W_{12} = 1/2 \nonumber \\
W_{n0} = p \quad  W_{n,n-1} = q/2  \quad  W_{n,n+1} = 1/2  \quad (n > 1)
%W_{20} = p \quad W_{21} = q/2 \quad W_{23} = 1/2 \nonumber \\
%W_{30} = p \quad W_{32} = q/2 \quad W_{34} = 1/2, \quad {\rm etc.}
\label{transitions}
\end{eqnarray}
Note that in many of these transitions the actual location of the rightmost 1 changes. We always define states in a frame moving with the front.

The equations for the probabilities are:
\begin{eqnarray}
P_0 W_{01} &=& \sum_{n=1}^\infty P_n W_{n0} \nonumber \\
P_1 (W_{10} + W_{12}) &=& P_0 W_{01} + P_2 W_{21} \nonumber \\
P_n (W_{n0} + W_{n,n-1} + W_{n,n+1}) &=& P_{n-1} W_{n-1,n} + P_{n+1} W_{n+1,n}   \quad (n > 1) \nonumber \\
%P_0 W_{01} &=& P_1 W_{10} +  P_2 W_{20} + P_3 W_{30} + P_4 W_{40} +...\nonumber \\
%P_1 (W_{10} + W_{12}) &=& P_0 W_{01} +  P_2 W_{21} \nonumber \\
%P_2 (W_{20} + W_{21} + W_{23}) &=& P_1 W_{12} +  P_3 W_{32} \nonumber \\
%P_3 (W_{30} + W_{32} + W_{34}) &=& P_2 W_{23} +  P_4 W_{43} \nonumber \\
%\ldots \nonumber \\
\end{eqnarray}

Using Eq.~(\ref{transitions}) we have:
\begin{eqnarray}
%\left({q \over 2}\right)  P_0 &=& \left({1+p\over 2} \right) P_1 + p (P_2 + P_3 + P_4 ...)  \nonumber \\
\left({q \over 2}\right)  P_0 &=& \left({1+p\over 2} \right) P_1 + p  \sum_{n=2}^\infty P_n \nonumber \\
\left({3 - q \over 2}\right)  P_1 &=& \left({q\over 2}\right) P_0 + \left({q\over 2}\right) P_2 \nonumber \\
\left({3 - q \over 2}\right)  P_{n} &=& \left({1\over 2}\right) P_{n-1} + \left({q\over 2}\right) P_{n+1} \quad (n > 1)
%\left({3 - q \over 2}\right)  P_2 & = & \left({1\over 2}\right) P_1 + \left({q\over 2}\right) P_3  \nonumber \\
%\left({3 - q \over 2}\right)  P_3 &=& \left({1\over 2}\right) P_2 + \left({q\over 2}\right) P_4
\label{transitions2}
\end{eqnarray}
%The structure of the equations for $P_2, P_3, ...$ are the same:
%\begin{equation}
%\label{pn}
%\left({3 - q \over 2}\right)  P_{n+1} = \left({1\over 2}\right) P_n + \left({q\over 2}\right) P_{n+2},
%\end{equation}
%for $n = 1, 2, 3, \ldots$.    

For $n > 1$, we make the ansatz
$P_{n} = a^{n-1} P_1$, and
inserting this in the last equation above, we find:
\begin{eqnarray}
\left({3 - q \over 2}\right) a &=& {1\over 2} + \left({q\over 2}\right) a^2 
\nonumber \\
a &=& { 3 - q  - \sqrt{9 - 10 q + q^2} \over 2q } = {1 \over 3} + {4 q \over 27} + {2 q^2 \over 243} \ldots. 
\label{aeq}
\end{eqnarray}

Next, we substitute $P_2 = a P_1$ into  the second line of Eq.~(\ref{transitions2}) and use
$\sum_{n=2}^\infty P_n = 1 - P_0 - P_1$, in the first line of Eq.~(\ref{transitions2}).
Solving these two equations we find:
\begin{eqnarray}
P_0  &=& {2 (1-q)(3 - (1+a)q) \over 6 - (5 + 2a) q + a q^2} \nonumber \\
&=& 1- {q \over 2}- {q^2 \over 12}- {11 q^3\over 216} - {137 q^3\over 3888} \ldots 
\end{eqnarray}
Thus
\begin{eqnarray}
P_1 &=&  {q \over 3}- {q^2 \over 54}- {19 q^3\over 972} +\ldots \nonumber \\
P_2 &=& a P_1 =  {q \over 9} + {7 q^2 \over 162} + {53 q^3\over 2916}+\ldots 
\nonumber \\
P_3 &=& a P_2 =  {q \over 27} + {5 q^2 \over 162} + {7 q^3\over 324} +\ldots 
\nonumber 
\end{eqnarray}

The velocity can be found from 
\begin{equation}
\label{velocity1}
 v = {1 \over 2} - {q P(\times01) \over 2}, 
\end{equation}
where $\times$ is any string of 0's and 1's, and we omit the zeros to the right.
In this case, $P(\times01) = P_1$, because  $|1\rangle$ 
is the only one-hole state that ends with $(01)$.  Thus, 
we have
\begin{equation}
\label{velocity2}
v = {1 \over 2} -  {q^2 \over 6} + {q^3 \over 108} + {19 q^4\over 1944} \ldots
\end{equation}

\begin{figure}
\begin{center}
\includegraphics[width=3.5in]{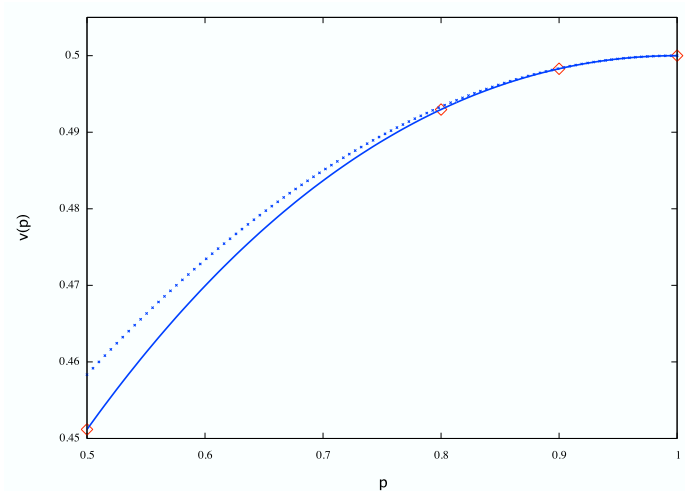}
\caption{Results for $v(p)$ in one dimension for $0.5 < p < 1$. The $\diamond$'s are the numerical results,  the dotted line is the quadratic approximation, and the solid line the power series of  Eq.~(\ref{velocity4}).}
\label{vnear1}
\end{center}
\end{figure}

We have precise numerical data for $v(q)$ for small $q$; see Figure \ref{vnear1}.  We find that Eq.~(\ref{velocity2}) is correct only up to quadratic order, as we might expect in the one-hole approximation.  For example, for 
$q = 0.1$, we find numerically that $v = 0.498292$, while $1/2 - q^2/6 = 0.49833$, so that the coefficient of  $q^3$ should be negative, not positive.  Note
$q^3 / 108 \approx 0.00001$. 

\subsection{Reduced distribution functions.}  
In order to go beyond quadratic terms in $q$, we introduce \emph{reduced distribution functions}.  This method would, in principle, allow the power-series expansion to be carried to arbitrary order. 

A reduced distribution function is probability to have a given pattern near the front for any pattern to the left. 
For example, as in Eq.~(\ref{velocity1}):
$$P(\times01) = \hbox{prob} (\ldots x x x 0 1 0 0 \ldots) $$
where the sites marked as $x$ are any string. Likewise, we define
$P(\times11)$, $P(\times001), P(\times101)$, and so forth.  Note that, for example: 
\begin{eqnarray}
P(\times001) + P(\times101) &=& P(\times01) \nonumber \\
P(\times011) + P(\times111) &=& P(\times11). \nonumber 
\end{eqnarray} 

We can derive a hierarchy of equations based on events that change the last $n$
sites.  For $n=2$, consider all events that change the probability that the
last two sites are (11).  We have:
\begin{eqnarray}
\left({p\over2}+ {p\over2}\right)P(\times001) &+& \left({1\over2}
+{1\over2}+{p\over2}\right) P(\times101)
  \nonumber \\
-\left({q\over2}+{q\over2}\right) P(\times011) &-&\left({q\over2}\right) P(\times111)= 0.
\label{11eqn}
 \end{eqnarray}
The positive terms represent events that increase the population of
states ending with (11), and the negative terms represent events that decrease that population.
Next we write all states in terms of  $P(\times01), P(\times001), P(\times011)$, i.e., states
that have a leading zero on the left.  For the other states, we use:
\begin{eqnarray}
P(\times101) &=& P(\times01) - P(\times001) \\
P(\times111) &=& P(\times11) - P(\times011) = 1 - P(\times01)  - P(\times011).
\end{eqnarray}
Then Eq.~(\ref{11eqn}) becomes:
\begin{equation}
{q \over 2} - {\left(3\over2\right)}P(\times01) +{\left(1+q\over2\right)}  P(\times001) + {\left(q\over2\right)}P(\times011) = 0.
\label{11eqn2}
\end{equation}
Now, we expect that $P(\times01) = {\cal O}(q), P(\times011) = {\cal O}(q),$ and $P(\times001) = {\cal O}(q^2)$, because a diffusive move
(weight $q/2$) is required to produce each empty site starting from state $|0\rangle$.  To order $q$ we find from Eq.~(\ref{11eqn2})
\begin{equation}
P(\times01) = {q \over 3} + {\cal O}(q^2)
\end{equation}
 which agrees with the leading behavior found in the one-hole approximation.  This implies that 
 the velocity is given by:
 \begin{equation}
\label{velocity3 }
v = {1 \over 2} - {q P(\times01) \over 2} =  {1 \over 2} - {q^2  \over 6} + {\cal O}(q^3)
\end{equation}
Note that there is no linear term, as mentioned above.
 
 For the next-order behavior, we use:
 \begin{eqnarray}  
P(\times011) & =& {q \over 9} +  {\cal O}(q^2) \label{16} \\
P(\times001) & =& {q^2 \over 9} +  {\cal O}(q^3)
\label{17}
\end{eqnarray} 
where the leading behavior in Eq.~(\ref{16}) is from the one-hole approximation. The second line follows from a simple argument: the leading behavior of
$P(\times001)$ is determined by $P(...1001)$, and its leading behavior is determined by the equation:
$$P(...1001) (3/2) = P(...101)(q/2) + P(...10001)(p/2 + q/2) + ... $$ 
Again $...$ represents a string of all 1's to the left. 
The second and higher-order
terms on the right-hand-side are of order $q^3$, so to leading order we find $P(...1001)  = (q/3) P(...101) = q^2/9$,
which proves Eq.~(\ref{17}).
 
Using these results, Eq.~(\ref{11eqn2}) implies
 $$P(\times01) = {q \over 3} + {2 q^2 \over 27} +  {\cal O}(q^3)$$
 which yields
 $$ v = {1 \over 2} - {q^2  \over 6}  - {q^3  \over 27} +  {\cal O}(q^4).$$
 For the case $q = 0.1$, these three terms give $v = 0.498296$, in close agreement
 with the numerical simulations, which give 0.498292.  
 
 We have carried this procedure to the next order, $n=3$, by straightforward extensions of what we have given above. The result for the velocity is:
 \begin{equation}
\label{velocity4}
v = {1 \over 2} - {q^2  \over 6} - {q^3  \over 27} - {49 q^4  \over 1215} + {\cal O}(q^5) 
\end{equation}
  
 For $q = 0.1$, the predicted velocity is now 0.4982923,
in complete agreement with the numerical result 0.498292. Even at $q = 0.5$, the prediction of Eq.~(\ref{velocity4}), $v = 0.4511831$, is within 0.2\% of the 
measured value, 0.45014. 

\section{Application to biology}
Our emphasis in this paper has been an analysis of the model introduced in the introduction. It is interesting, nevertheless, to make some comments on the relationship of this model to real biological systems. Needless to say, our view of wound repair is very much oversimplified. In a real tissue there are various types of cells such as stem cells which have different behavior with respect to proliferation than others. Further, the proliferation cycle is complex, and involves time delays that we have not considered except in a rough way. Also, the initiation of wound healing is probably mediated by chemical signals rather than cell proximity as we have assumed \cite{walker04a}. 

However, if we are interested in macroscopic features such as the velocity and shape of the moving front, we are entitled to hope that many of these details will be unimportant. We can then ask how to translate the parameters of our model to a real system. We will take as an example the experiment of Sheardown and Cheng \cite{sheardown96} on the wounding of rabbit corneas.   

In \cite{sheardown96} the emphasis was on modeling with the FK equation. To this end the authors measured $D$ in Eq.~(\ref{FKeqn}) by looking at the initial stage of invasion  of cultured cells, and found $D = 1.61 \cdot 10^{-6} {\rm mm^2/s}$. The parameter $k$ in  Eq.~(\ref{FKeqn}) is related to the mitotic rate of cells which was measured by labeling with a dye:  $k^{-1} = 4.3 \quad {\rm days}$. Using these parameters the authors found reasonable agreement for  the velocity of the front. Further, for these parameters, the shape of the front is quite `fuzzy', that is, the width, $w$, is many cells across so that the wound fills in gradually, as observed. We should note that this is in sharp contrast to other observations \cite{sheratt02,maini04} where the advancing front is quite sharp. We will return to this point below.

We now attempt to translate these observations into the parameters of our discrete model. We need to define units of length and time. For length it is natural to take a typical cell size, $ d = 10\mu$ as the lattice unit. It is clear that we can define the \emph{hopping time}, $\tau_{hop}$ by  $D = d^2/2\tau_{hop}$. This turns out to be about 108  seconds for the rabbit cornea. However, there is another characteristic time, the cell \emph{cycling time}, $\tau_{cyc} =1/k$. This is $3.7 \cdot 10^5$ seconds for the same experiment. In our model, in $N$ computer cycles there are $Np$ cell cycles and $Nq$ hops.  Thus our time unit should be $q\tau_{hop} + p\tau_{cyc}$.

To determine the biological $p$ note that $p/q = \tau_{cyc}/\tau_{hop}$.
For the rabbit experiment we get $p=3 \cdot 10^{-4}$. This is the regime of very diffuse, fuzzy interfaces, as observed. In this regime FK modeling should be reasonable, though, as Figure~\ref{v(p)} shows, there are still differences between FK and the discrete model in this regime.  

If we apply the same set of considerations to the systems studied in \cite{sheratt02,maini04} we find a contradiction. The width of the interface should be quite substantial for the small $p$'s relevant to biological experiments, cf. Figure~\ref{w(p)}. In fact, FK modeling shows the same thing. However, direct observation in these cases shows that the front is quite sharp.

A possible solution to this quandary is given by \cite{walker04a} where it is pointed out that cell-cell adhesion can have an effect on wound healing in some systems. In fact, for the cells that they study  they can regulate the adhesion, and hence the front width, by controlling the supply of Ca$^{++}$ ion in the solution bathing the cells. A detailed discrete model in their paper shows these effects too. This is an interesting avenue for future work.  

\section{Summary and conclusions}
In this paper we have extended the work of \cite{kerstein86,bramson86} on a discrete model. We have shown that the model can be interpreted as a representation of the important biological process of wound healing. We have given numerical results in one and two dimensions, and a power-series expansion of the velocity around $q=0$ in one dimension. We have shown that the biologically interesting regime is that of $p\ll1$.

There are a number of further extensions of this work that could be pursued. Our method of reduced distribution functions should be applicable to models with more complex rules as long as a sensible expansion parameter, analogous to $q$, is present. We do not understand why the convergence to the FK limit is different in our case than in the generic cases discussed in \cite{moro03}. A more extensive numerical study may be called for. 

We think that the most interesting extension of the model would be to include cell-cell adhesion, in the spirit of \cite{walker04,walker04a}. This work is in progress. 

\paragraph{Acknowledgements} LMS would like to thank P. Maini and J. Devita for helpful discussions. Supported in part by NSF grant DMS-0244419. EK is supported by NIH  grant  CA085139-01A2.

\bibliography{woundheal}
\end{document}